\documentclass[fleqn,usenatbib]{mnras}

\usepackage{newtxtext,newtxmath}

\usepackage[T1]{fontenc}

\DeclareRobustCommand{\VAN}[3]{#2}
\let\VANthebibliography\thebibliography
\def\thebibliography{\DeclareRobustCommand{\VAN}[3]{##3}\VANthebibliography}

\usepackage{graphicx}
\usepackage{amsmath}
\usepackage{orcidlink}

\title[Sub-Solar Planetesimals]{Planetesimal Formation Across the Stellar Mass Spectrum and its Influence on Exoplanet-Inherited Volatile Budgets}

\author[J. Williams et al.]{
Joe Williams,$^{1}$\thanks{E-mail: jw1436@exeter.ac.uk}\orcidlink{0009-0008-8176-1974}
Sebastiaan Krijt,$^{1}$\orcidlink{0000-0002-3291-6887}
Joanna Dr{\k{a}}{\.z}kowska,$^{2}$\orcidlink{0000-0002-9128-0305}
and Tim Lichtenberg$^{3}$\orcidlink{0000-0002-3286-7683}
\\
$^{1}$Department of Physics and Astronomy, University of Exeter, Stocker Road, Exeter EX4 4QL, UK\\
$^{2}$Max Planck Institute for Solar System Research, Justus-von-Liebig-Weg 3, 37077 Göttingen, Germany\\
$^{3}$Kapteyn Astronomical Institute, University of Groningen, Groningen, Netherlands\\
}

\date{Accepted XXX. Received YYY; in original form ZZZ}

\pubyear{2026}

\begin{document}
\label{firstpage}
\pagerange{\pageref{firstpage}--\pageref{lastpage}}
\maketitle

\begin{abstract}

Protoplanetary discs emerging from collapsing molecular clouds are capable of forming planetesimals at the water snowline during both the cloud collapse and Class II disc phases; such a scenario could be responsible for creating the carbonaceous/non-carbonaceous (CC/NC) heterogeneity observed in the Solar System, and bears important implications for emergent planetary compositions. We use 1D simulations of a viscously evolving disc coupled with cloud collapse and planetesimal formation to explore how planetesimal formation during disc build-up varies across the stellar mass spectrum. We find a keen sensitivity of planetesimal formation timing, location, and outcomes on stellar mass. Discs around all investigated stellar masses form planetesimals in the Class II phase, but only the disc around low-mass M-dwarfs ($M_\star=0.1M_\odot$) fails to form them during the infall phase. There is also a clear chemical heterogeneity in planetesimal populations (water-wet and dry) in discs born from clouds of $M_{\rm{cloud}}\geq0.3M_\odot$. Discs around low-mass M-dwarfs form and undergo extremely fast pebble drift ($t<2$ Myr), forming planetesimals well within the half-life of Aluminium-26. This leads to dehydrated planetesimals in all M-dwarf disc formation cases considered. We argue that the variation in disc evolution across stellar mass makes it hard to pinpoint a common $t=0$ for all discs, and that exoplanets emerging from dehydrated planetesimals around low-mass M-dwarfs will be born volatile-poor - potentially explaining the lack of rocky world atmospheres seen by JWST.
\end{abstract}


\section{Introduction}

Planet formation is expected to start earlier than previously thought in protoplanetary disc evolution, which may require the formation of planetesimals at early stages \citep{manara2018_mass_budget, drazkowska_pp7_2023}. Studies of planet and planetesimal formation often rely on the streaming instability \citep{youdin&goodman2005_SI, johansen2007_SI} to create these planetesimals, although the criteria for this instability can be met in certain conditions. Such conditions often require dust traps and pressure bumps \citep[e.g.][]{lau2022_core_bumps, lau2024_sequential_planets, jiang&ormel2023_ring_planets, sandor2024_plts_in_traps}, the water snowline \citep[e.g.][]{saito&sirono2011_iceline_plts, ida&guillot2016_iceline_plts, schoonenberg&ormel2017_iceline_plts, drazkowska&alibert2017_snowline_plts}, material infall \citep{zhao2025_plts_by_infall}, or some combination thereof \citep{drazkowska&dullemond2018_disk_buildup, lichtenberg2021_bifurcation}.

\citet[][hereafter DD18]{drazkowska&dullemond2018_disk_buildup} studied the combined effects of material infall as a means of protoplanetary disc build-up and water condensation on planetesimal formation, based upon the model by \citet{drazkowska&alibert2017_snowline_plts}. The combined effect of a dust ``traffic jam'' resulting from a changing fragmentation velocity across the water snowline and the ``cold-finger effect'' from the diffusion of water vapour across the snowline was sufficient to form planetesimals. They found, however, that it is easier to form planetesimals in the Class II phase compared to the disc build-up phase.

The models of DD18 were expanded on by \citet[][hereafter L21]{lichtenberg2021_bifurcation} to explore possible evolutionary pathways of the young Solar system. L21 explored disc build-up via a collapsing molecular cloud as a means to create the observed spatial dichotomy between carbonaceous chondrites (CC) and non-carbonaceous chondrites (NC) in the solar system. These chondrites contain Calcium-Aluminium-rich Inclusions (CAIs) that are expected to have formed alongside the young sun \citep{connelly2012_CAI}, possibly in the hot, inner regions of the protosolar disc \citep{woitke2024_CAIs}. The early formation of CAIs mean they act as an effective means to date the earliest stages of planet formation in the solar system. The origin of the CC/NC dichotomy is hotly debated \citep[see e.g.][and references therein]{kruijer2017_jupiter_CCNC, desch2018_jupiter_CCNC, kleine2020_CCNC, colmenares2024_ccnc_processing, marty2024_cometsIII}, but we address one of the pillars of L21's argument: Aluminium-26 ($^{26}$Al).

$^{26}$Al may have played an important role in creating the Solar dichotomy (L21). It is expected to be the dominant planetesimal heating mechanism for Solar-like abundances \citep{nittler&ciesla2016_extraterrestrial, lugaro2018_al26_variability, parker2020_planet_birth}, and may have been injected into the Solar nebula via nearby supernovae \citep[][]{desch2018_jupiter_CCNC, nanne2019_CCNC_origin, jacquet2019_meteorite_anomalies, eatson2024_al26_injection} or AGB and Wolf-Rayet stars \citep{lugaro2018_al26_variability, lichtenberg2023_PP7, parker&schoettler2023_enrichment_by_agb}. The injected $^{26}$Al became incorporated into first-generation planetesimals formed at $t< \tau_{^{26}\rm{Al}}$ during the cloud collapse phase, where $\tau_{^{26}\rm{Al}}$ designates the mean lifetime of $^{26}$Al. Planetesimals forming later than $\tau_{^{26}\rm{Al}}$ experience significantly decreased internal heating devolatilisation. This dust was spread outward by the viscous expansion of the disc, separating the CC-rich material from the $^{26}$Al-deprived dust that was accreted later. Both the disc spreading and $^{26}$Al-deprived second-generated planetesimals then created the temporal and spatial distinction between CC and NC.

Exoplanets are indirectly influenced by $^{26}$Al heating through the volatile contents of their constituent planetesimals. Exoplanetary systems' initial $^{26}$Al abundances may vary considerably \citep{lugaro2018_al26_variability}, and it may not be common for Solar-like stellar nebulae to be injected with $^{26}$Al \citep{parker2014_al26_solar_enrichment}. Despite this, the presence of $^{26}$Al will play an important role in setting exoplanet volatile budgets. We therefore aim to explore planetesimal formation around different stellar masses. Do all stars form planetesimals in the same way and the same time, and do their emergent exoplanets therefore have the same compositions  \citep{krijt_pp7_2023}? How important is stellar and parent cloud mass in determining planet formation outcomes?

In addition, the large sample of recent JWST data from both the JDISCS \citep{arulanantham2025_JDISCS} and MINDS programs \citep{kamp2023_MINDS, henning2024_MINDS_overview} covers a large range of stellar and protoplanetary disc parameters, including M-dwarf stars \citep{long2025_30myr_disc, arabhavi2025_minds_VLMS}. The extensive study of discs with varying properties around a diverse stellar population further motivates our investigation into planetesimal formation and possible pebble drift histories around other types of star \citep[e.g.][]{mah2023_vlms_C/O, andama2024_planetesimal_low_metal}. 

Our work is structured as follows. We outline our methodology and initial conditions in Section~\ref{sec:methodology}. We present the planetesimal formation outcomes, the evolution of mass reservoirs, and the relation between $\dot{M}_\star$ and $M_\star$ compared against observations in Section~\ref{sec:results}. We discuss the ramifications for planet formation, particularly around very low mass stars, as well as disc evolution implications in Section~\ref{sec:discussion}. Finally, we conclude our investigation in Section~\ref{sec:conclusions}.

\section{Methodology}
\label{sec:methodology}

\subsection{Disc and Stellar Model}
\label{subsec:disc model}
We use the model from DD18\footnote{https://github.com/astrojoanna/DD-diskevol}, which couples the planetesimal formation scheme of \citet{drazkowska&alibert2017_snowline_plts} with the cloud collapse model of \citet{shu1977_collapse, ulrich1976_rotating_cloud}. The cloud, assumed to be well represented by an singular isothermal sphere, begins to collapse at $t=0$ and infalls at a constant rate of $\dot{M}=0.975 c_s^3/G$ for sound speed $c_s$ before abruptly stopping after the cloud has collapsed into the star and disc. The dust is evolved according to the two-population algorithm \citep{birnstiel2012_twopoppy}, although we do not include the growth of dust in the molecular cloud phase, where the growth timescale is expected to be long \citep{ormel2009_growth_in_clouds} and grains are unlikely to grow beyond $\sim10\mu$m \citep{lombart2026_dust_in_envelopes}.

We also model the transport of water ice, with a sublimation and condensation scheme following \citet{ciesla&cuzzi2006_water_evol}. The fragmentation velocity of silicates and particles covered in water ice is set to 1 ms$^{-1}$ and 10 ms$^{-1}$ respectively \citep[e.g.][]{blum&wurm2008, wada2013}, allowing for a traffic jam to occur at the water snowline. Additionally, water vapour released inside the snowline can diffuse back across the snowline, following concentration gradients \citep{stevenson&lunine1988_recondense, cuzzi&zahnle2004}; this is the so-called ``cold-finger effect''. Both the traffic jam and cold-finger effect are required to form planetesimals at the snowline via the streaming instability, with sufficiently low turbulence \citep[][DD18]{drazkowska&alibert2017_snowline_plts}. The planetesimals form at a rate $\dot{\Sigma}_{\rm{plts}}$ as prescribed by \citet{drazkowska2016, drazkowska&alibert2017_snowline_plts}:

\begin{equation}
    \dot{\Sigma}_{\text{plts}} = \zeta \Sigma_{\text{dust}}(\rm{St}>10^{-2})\Omega_{K}
\end{equation}

\noindent where $\zeta$ is the planetesimal formation efficiency (set to $10^{-3}$ here) and $\rm{St}$ the Stokes number of dust undergoing collapse. We refer the interested reader to the aforementioned papers for further details on the model used here.

We use $\alpha_{\rm{visc}}=10^{-3}$ and $\alpha_{\rm{turb}}=10^{-5}$ following DD18 and L21, which are close to and considerably below the upper limits inferred from observations \citep[$\alpha_{\rm{visc}}\lesssim 3\times10^{-3}$ and $\alpha_{\rm{turb}} \lesssim 10^{-3}$][]{rosotti2023_turbulence}. These describe the viscous gas evolution ($\alpha_{\rm{visc}}$) and turbulence dictating the fragmentation-limited dust size in the collision-fragmentation cascade ($\alpha_{\rm{turb}}$). Selecting $\alpha_{\rm{turb}}<\alpha_{\rm{visc}}$ is motivated by observed quiescent midplanes (\citealt{turner2014_PP6_turbulence}; DD18; \citealt{rosotti2023_turbulence}) and other planetesimal formation simulations \citep{carrera2017_plts_formation_photoevap, ercolano2017_plts_formation_xray}. We refer the reader to DD18 for a thorough study of $\alpha$ in the context of their model.

To capture the effects of different stellar types, we change the stellar mass, radius, and luminosity between each simulation. We calculate static luminosities based on models by \citet{baraffe2015_luminosities}, picking the radius and effective temperature of the star at 10 Myr; we choose this evolutionary stage as representative stellar properties for the early stellar evolution, also following previous simulations (DD18, L21). We pick 1$M_\odot$ star has $T_{\rm{eff}}=4363$ K and $R_\star=1.189R_\odot$, leading to $L_\star=0.45L_\odot$. This differs from the choice of $T_{\rm{eff}}=5800$ K, $R_\star=R_\odot$, and $L_\star=L_\odot$ in L21, but the viscous heating due to infalling material dominates the infall-stage planetesimal formation; as such, the cloud mass and rotation rate are the most influential on planetesimal formation. We compare our choice against updated results from L21 in Section~\ref{appendix:comparison to L21}, and discuss the implications of our choices surrounding stellar properties in Section~\ref{subsubsec:stellar evolution}.

\subsection{Cloud Initial Conditions}
\label{subsec:cloud initial conditions}

\begin{table}
	\centering
	\caption{Parameters of the baseline models presented in this paper. Each model is run with $\alpha_{\rm{visc}}=10^{-3}$ and $\alpha_{\rm{turb}}=10^{-5}$, and a dust-to-gas ratio of 0.01.}
	\label{tab:model parameters}
	\begin{tabular}{cccc} 
		\hline
		$M_{\rm{cloud}}$ ($M_\odot$) & $\Omega_{\rm{cloud}}$ ($10^{-15}$ rad s$^{-1}$) & $T_{\rm{eff}}$ (K) & $R_{\star}$ ($R_{\odot}$) \\
		\hline
        \hline
		0.1 & 22.0 & 3050 & 0.42 \\
		0.3 & 12.8 & 3363 & 0.71 \\
		0.5 & 9.90 & 3651 & 0.88 \\
        0.8 & 7.83 & 4041 & 1.06 \\
        1.0 & 7.00 & 4363 & 1.19 \\
        1.4 & 5.92 & 5392 & 1.63 \\
		\hline
	\end{tabular}
\end{table}

To create varied initial conditions for star and disc formation, we begin with the mass $M_{\rm{cloud}}$, radius $R_{\rm{cloud}}$ and rotation rate $\Omega_{\rm{cloud}}$ used in DD18 ($M_{\rm{cloud}}=1.0M_\odot$, $\Omega_{\rm{cloud}}=7\times10^{-15}~\rm{rad~s}^{-1}$). We then select different cloud masses given in Table~\ref{tab:model parameters} and calculate the cloud radius according to \citet{dullemond2006b_core_properties}:

\begin{equation}\label{eqn:cloud radius}
    R_{\rm{cloud}}=\dfrac{GM_{\rm{cloud}}}{2c_s^2},
\end{equation}

\noindent where $c_s=\sqrt{k_B T/\mu m_p}$ is the speed of sound in the isothermal cloud. We choose to set $T=10$K. When the cloud collapses, it deposits material inside its centrifugal radius, which increases with time until it reaches the cloud's centrifugal radius:

\begin{equation}\label{eqn:centrifugal radius}
    R_{\rm{centr}}=\dfrac{\Omega_{\rm{cloud}}^2 R_{\rm{cloud}}^4}{GM_{\rm{cloud}}} = \dfrac{\Omega_{\rm{cloud}}^2G^3 M_{\rm{cloud}}^3}{16c_s^8}.
\end{equation}

\noindent Since $R_{\rm{centr}}\propto M_{\rm{cloud}}^3$, more massive clouds will create physically larger discs \citep{dullemond2006b_core_properties}.

\begin{figure}
	\includegraphics[width=1\columnwidth]{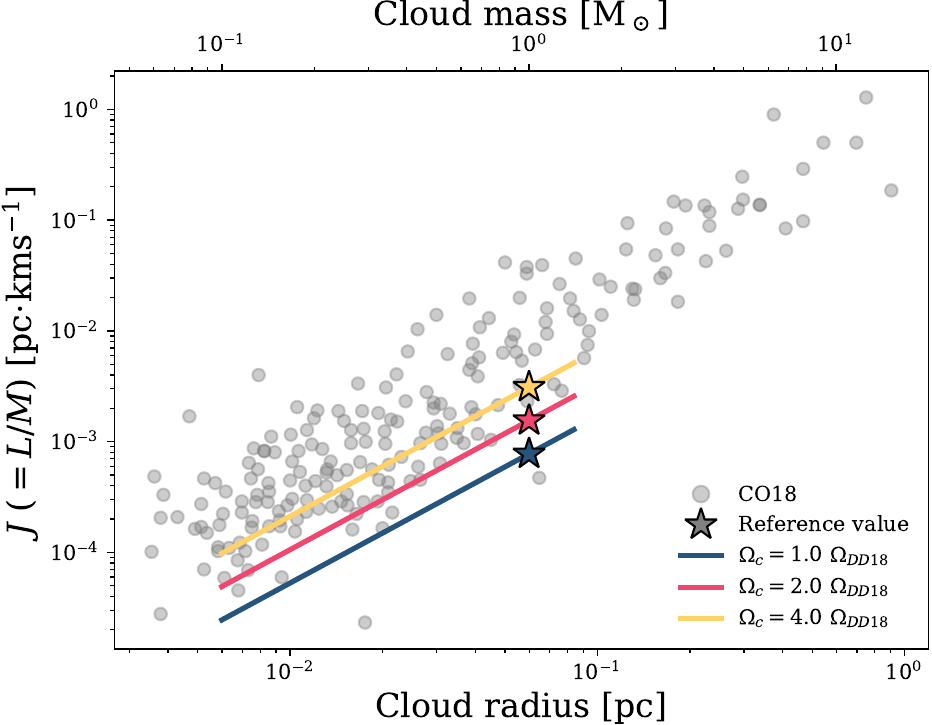}
    \caption{Specific angular momentum $J=L/M$ of molecular clouds from observational data with the initial conditions used in this work (lines) plotted atop. The observational data was adapted from Fig.~11 of \citet{chen&ostriker2018_cloud_J} and references therein. The `reference value' denotes the cloud mass and $J$ used to construct the initial conditions for constant $\beta$. The reference value for the blue line uses $M_{\rm{cloud}}$ and $\Omega_{\rm{cloud}}$ from \citet{drazkowska&dullemond2018_disk_buildup}, which we increase for other initial conditions (red and yellow lines).}
    \label{fig:cloud initial conditions}
\end{figure}

To simulate different systems, we scale $\Omega_{\rm{cloud}}$ so that the cloud's angular momentum and radius follow the relation $J\propto R^{1.5}$ \citep[see, e.g.][]{chen&ostriker2018_cloud_J}. We therefore express $\Omega_{\rm{cloud}}$ using a reference value $\Omega_0$ and $R_0$:

\begin{equation}\label{eqn:Omega rescaling}
    \Omega_{\rm{cloud}} = \Omega_0 \left( \dfrac{R_{\rm{cloud}}}{R_0} \right)^{-0.5},
\end{equation}

\noindent where $\Omega_{\rm{0}}=7\times 10^{5}$ rad s$^{-1}$ and  $R_{\rm{0}}\approx12400$ au are the reference values (DD18). We derive this equation in Appendix~\ref{appendix:omega scaling derivation}. We created a set of initial conditions that maintain a constant ratio between the rotational and gravitational potential energy of the sphere \citep[i.e. $\beta$, see e.g. Appendix B1 of][]{bate2018_disc_statistics} using this equation.

We investigate two additional sets of initial conditions, where each set uses a different value of $\Omega_{\rm{0}}$. The first set uses $\Omega_{\rm{0}}=7\times10^{-15}$ rad s$^{-1}=\Omega_{\rm{DD18}}$; the second uses $\Omega_{\rm{0}}=2\Omega_{\rm{DD18}}$; and the third uses $\Omega_{\rm{0}}=4\Omega_{\rm{DD18}}$. We demonstrate that these initial conditions follow the J-R relation in Fig.~\ref{fig:cloud initial conditions}, but leave discussion of varying $\Omega_0$ to Section~\ref{subsec:variable cloud parameters}.

Finally, we also verify that the simulations follow observational trends by checking the relationship between stellar mass accretion rate and stellar mass as a function of time. We discuss this in detail in Section~\ref{subsec:mdot-m relation}.

\section{Results}
\label{sec:results}

\subsection{Snowline Migration}
\label{subsec:snowline migration}

We firstly focus on the evolution of the snowline, since this dictates the dynamics of planetesimal formation; the thermal histories of discs born from collapsing clouds changes significantly depending on the parent cloud parameters.

\begin{figure}
	\includegraphics[width=1\columnwidth]{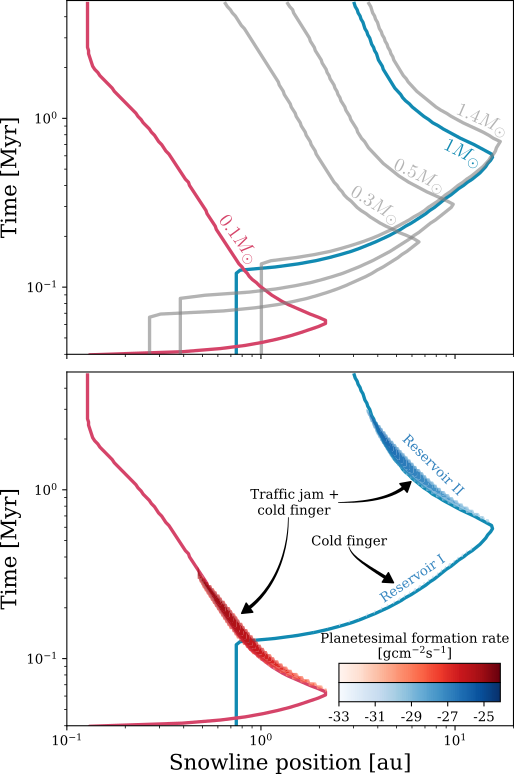}
    \caption{\textit{Top:} Position of the water snowline as a function of time for different cloud masses (labelled). The 0.1$M_\odot$ and 1$M_\odot$ cases are highlighted as the focus of our results. The $0.8M_\odot$ simulation is omitted for visual clarity. \textit{Bottom}: Water snowline position with the planetesimal formation rate overlaid in different colours (red for $0.1M_\odot$ and blue for $1M_\odot$). The $1M_\odot$ cloud features two distinct reservoirs of planetesimals, whereas the $0.1M_\odot$ case only forms planetesimals due to the cold finger effect and traffic jam. The snowlines are dashed lines in the bottom panel to distinguish from formation rates. The colourbar shows $\ln(\dot{\Sigma}_{\rm{plts}}/\rm{g~cm^{-2}~s^{-1}})$.}
    \label{fig:snowline position}
\end{figure}

More massive clouds, which produce more massive discs, experience infall later and undergo significantly more viscous heating, as can be seen in Fig.~\ref{fig:snowline position}. The $0.1M_\odot$ cloud sees its snowline extend up to $\sim$1 au by 0.06 Myr, whereas the $1.4M_\odot$ cloud deposits considerably more material into the disc; the subsequent viscous heating pushes the snowline to almost 20 au by $\sim$1 Myr. The snowlines migrate at different rates depending on the parent cloud properties. The $0.1M_\odot$ simulation sees its snowline migrate outwards at an average speed of $\sim$33 au Myr$^{-1}$, whereas the snowline in the $1M_\odot$ simulation recedes at an average speed of $\sim$41 au Myr$^{-1}$. Less massive clouds end their infall sooner, meaning the $0.1M_\odot$ cloud experiences much less viscous heating than the more massive counterparts, leading to the snowline migrating much less, but still at a comparable speed to higher mass systems.


Discs thermally and viscously evolve on much longer timescales when they originate from more massive clouds. Only the $0.1M_\odot$ system undergoes infall and accretes a large portion of its gas disc within 5 Myr. In all models, the snowline position is set by solely the stellar luminosity after the viscous heating becomes negligible (2 Myr for $0.1M_\odot$ and $>5$ Myr for $M_{\rm{cloud}}\geq0.3 M_\odot$). The evolutionary timescales and available mass reservoirs from the collapsing cloud set the stage for the subsequent planetesimal formation at the snowline.

\subsection{Planetesimal Formation}
\label{subsec:planetesimal formation}

Planetesimal formation at the snowline is sensitive to the thermal history of the disc, the migration of the snowline, and the material delivered to the snowline; we focus on the $0.1M_\odot$ and $1M_\odot$ simulations in the bottom panel of Fig.~\ref{fig:snowline position} as representative cases of different disc histories.

Firstly, we note that the planetesimal formation rate appears dissimilar to the original results from DD18 and L21; this was corrected in \citet{drazkowska&dullemond_2023_corrigendum}, although the net planetesimal mass and formation rates in the aforementioned works remain unchanged. Here, we use the updated model, and find that the belt of formed planetesimals is more narrow, particularly in Reservoir II. We also compare our results to those of L21 in Appendix~\ref{appendix:comparison to L21}, as we use slightly different parameters ($T_{\rm{eff}}=4363~\rm{K}$, $R_\star=1.189R_\odot$, and $L_\star=0.45L_\odot$, whereas L21 use $T_{\rm{eff}}=5800~\rm{K}$, $R_\star=R_\odot$, and $L_\star=L_\odot$). The only key differences is the snowline position before the infall phase, whereas the systems evolve very similarly otherwise.

We define two planetesimal reservoirs, ``Reservoir I'' and ``Reservoir II'', following L21; these reservoirs are formed from distinct episodes of formation at different orbital radii and time periods. For the solar-mass cloud, Reservoir I forms during the infall stage followed by Reservoir II in the Class II phase after a brief pause in planetesimal formation. We therefore define Reservoir I to be formed during the infall stage due to the cold-finger effect, and Reservoir II in the Class II stage from both the cold-finger effect and a traffic jam of pebbles \citep[][DD18, L21]{drazkowska&alibert2017_snowline_plts}. However, if there is no pause in planetesimal formation, these reservoirs may not be distinct in reality. We nonetheless distinguish these reservoirs as I and II in this work to highlight the mechanism and timing of planetesimal formation. We expand on the formation of potentially distinct reservoirs in Sections~\ref{subsec:mdwarf planets} and ~\ref{subsec:variable cloud parameters}.

The disc in the $1M_\odot$ simulation experiences planetesimal formation during the infall stage when the cloud is collapsing, whereas the disc in the $0.1M_\odot$ one does not, most likely due to a lower pebble mass flux to the snowline. This creates planetesimal Reservoir I. This is labelled in Fig.~\ref{fig:snowline position}, although it is difficult to visually distinguish planetesimal formation from the line marking the snowline; this is due to the relatively low formation rates compared to the Class II stage, and the low mass produced in Reservoir I. Additionally, we note that Reservoir I does not form ubiquitously across the stellar mass spectrum - only the disc around the $0.1M_\odot$ star does not form planetesimals during cloud collapse, whereas all other discs do.

The dependence of planetesimal formation on stellar and cloud mass can be clearly seen between the panels in Fig.~\ref{fig:mass reservoirs}. The disc in the low mass cloud ($0.1M_\odot$) forms extremely quickly ($<100$ kyr) and continuously forms planetesimals in the Class II stage until the dust reservoir is mostly used up. The disc in the higher mass cloud ($1M_\odot$), however, produces planetesimals both during the infall and Class II stage with a break in between. This pause occurs when the snowline is pushed far from the central star, with formation resuming only during the Class II stage (DD18, L21). This creates a clear \textit{temporal} distinction between Reservoir II. This pause in planetesimal formation can also be seen for different stellar masses and $\Omega_{\rm{cloud}}$ in Fig.~\ref{fig:planetesimal mass}.

In the intermediate $0.5M_\odot$ case, however, there is a kink in planetesimal formation as the simulation transitions from the infall stage to the Class II stage. Whilst these populations are not formed in temporally distinct episodes like the $1M_\odot$ case, the planetesimals will still be distinct in terms of their volatile content. We discuss this further in Section~\ref{subsubsec:planetesimal reservoirs}.

\subsection{Mass Reservoir Evolution}
\label{subsec:mass reservoirs}

\begin{figure*}
	\includegraphics[width=1\textwidth]{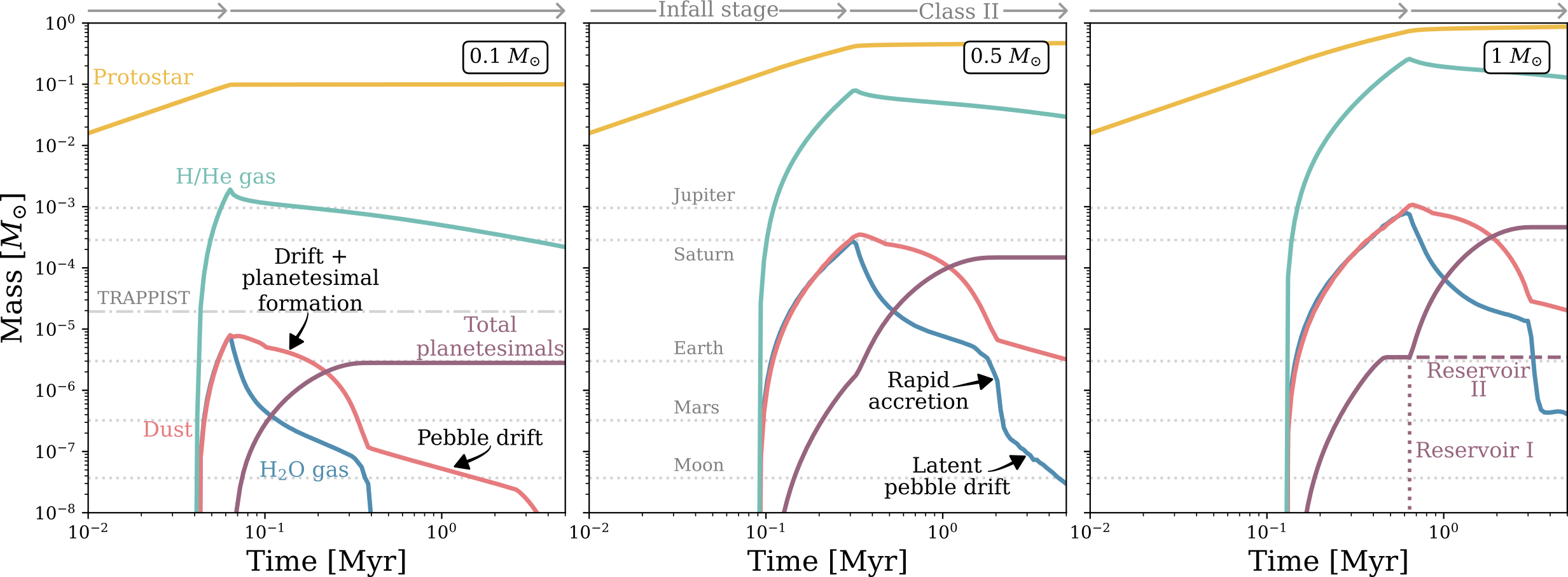}
    \caption{Evolution of mass reservoirs as a function of time. Each panel shows a different simulation with different cloud masses ($0.1M_\odot$, $0.5M_\odot$, and $1M_\odot$, labelled). Each mass reservoir is labelled with key physics mechanisms labelled. The mass of solar system planets are labelled, and the total mass of the planets in the TRAPPIST-1 system is shown in the left panel.}
    \label{fig:mass reservoirs}
\end{figure*}

Each simulation initially evolves in a similar way during cloud collapse, before quickly diverging depending on the cloud properties. Less massive clouds experience collapse more rapidly and material is deposited closer to their star compared to their higher mass counterparts; this leads to different timescales for disc evolution. We elucidate on this here by exploring the evolution of different mass reservoirs, which are presented in Fig.~\ref{fig:mass reservoirs}.

\subsubsection{Growth of the Central Star}
\label{subsubsec:stellar mass}

The collapsing cloud feeds both the central star and the gas and dust disc, with the vast majority of the cloud material ending up in the central star. Smaller mass clouds see practically all of their material ending up in the central star, whereas larger mass clouds produce substantial discs. The $1.4M_\odot$ cloud builds a $1M_\odot$ star by 5 Myr, which continues to accrete. This can be seen in the yellow protostar and green H/He gas curves in Fig.~\ref{fig:mass reservoirs}, where all of the mass reservoirs are labelled alongside relevant processes.

\subsubsection{Dust and Gas Reservoirs}
\label{subsubsec:dust reservoirs}

Less massive clouds deposit less material into the forming protoplanetary disc, which is also deposited closer to the star due to the cloud's smaller centrifugal radius (see equation~\ref{eqn:centrifugal radius}). The disc formed from the $0.1M_\odot$ cloud contains orders of magnitude less dust mass than those in the $0.5M_\odot$ and $1M_\odot$ clouds (see Fig.~\ref{fig:mass reservoirs}): the same disc contains just over $1M_{\rm{Jup}}$ of gas and $1M_\oplus$ of dust. Conversely, more massive clouds produce more massive discs: the discs emerging from the $0.5M_\odot$ and $1M_\odot$ clouds are supplied with $\sim1M_{\rm{Sat}}$ of dust and nearly $0.1M_\odot$ of gas, and $1M_{\rm{Jup}}$ of dust and $\sim0.2M_\odot$ of gas respectively. We also show the spatial and temporal evolution of the dust reservoir for the same clouds shown in Fig.~\ref{fig:mass reservoirs} in Appendix~\ref{appendix:dust maps}, which shows where the dust lies with respect to the snowline.

The mass of the cloud, and therefore the mass and size of the disc, as well as the stellar mass, strongly influences the overall evolutionary timescale. The dust reservoir is depleted much more quickly than the gas one in the $0.1M_\odot$ cloud; this is due to the shorter pebble drift timescales around lower mass stars and brown dwarfs \citep{pinilla2022_vlms_drift} and the rapid creation of planetesimals post-infall. Through both pebble drift and planetesimal formation, the $0.1M_\odot$ cloud rapidly depletes its dust reservoir to 1\% of its maximum value in $\sim0.4$ Myr. Planetesimal formation in the Class II phase acts as an additional sink for dust and ice on top of pebble drift. This is clear from the fact that the gradient of the dust mass over time suddenly becomes shallower when the planetesimal mass plateaus - that is, halting planetesimal formation leads to only pebble drift depleting the dust reservoir, incurring a lower rate of dust mass loss. We label in Fig.~\ref{fig:mass reservoirs} the relevant dust loss mechanisms to rapid dust evolution in the disc from the $0.1M_\odot$ cloud. 

We can conclude from this that more massive discs are more likely to be formed around more massive stars, having been born from more massive molecular cloud, although a plethora of discs around very low mass stars have been observed \citep{arabhavi2025_minds_VLMS, long2025_30myr_disc}. Work by \citet{bate2018_disc_statistics} also indicates that higher mass stars tend to have lower disc-to-star mass ratios, contrary to our findings here, but we attribute this to the limitations of a 1D model instead of a full hydrodynamical code; statistical properties of disc formation are therefore best left to simulations tailored to the dynamics of disc formation.

\subsubsection{Evolution of the Disc's Water Reservoir}

For each simulation in Fig.~\ref{fig:mass reservoirs}, the total (i.e. radially- and vertically-integrated) water vapour mass in the disc reservoir increases during the infall stage before rapidly decreasing with the formation of planetesimals. The water then undergoes an extremely rapid mass loss phase at $\sim0.4$ Myr for the $0.1M_\odot$ cloud, occurring later (2-3 Myr) for the higher mass clouds. This rapid mass loss coincides with a reduction in pebble flux to the water snowline as most of the material has already drifted - this is also why planetesimal formation shuts off (see Section~\ref{subsec:pebble drift} for further details on pebble drift). The same process occurs for the $0.5M_\odot$ and $1M_\odot$ clouds. We note that our water vapour mass and depletion timescale is comparable to previous pebble drift simulations \citep[e.g.][]{houge2025_smuggled_water}.

Pebbles that take longer to grow and drift from the distant regions of the disc then supply a trickle of water ice, leading to a shallower gradient that can be seen after the rapid water loss in the form of latent pebble drift \citep[][]{kalyaan2021, houge2025_smuggled_water}. Water is accreted onto the $0.1M_\odot$ star extremely rapidly: the viscous timescales at $\lesssim1$ au are much shorter than 0.1 Myr for $\alpha_{\rm{visc}}=10^{-3}$. The different viscous timescales between stellar masses lead to different subsequent evolution after the rapid accretion phase.

\section{Discussion}
\label{sec:discussion}

\begin{figure*}
	\includegraphics[width=1\textwidth]{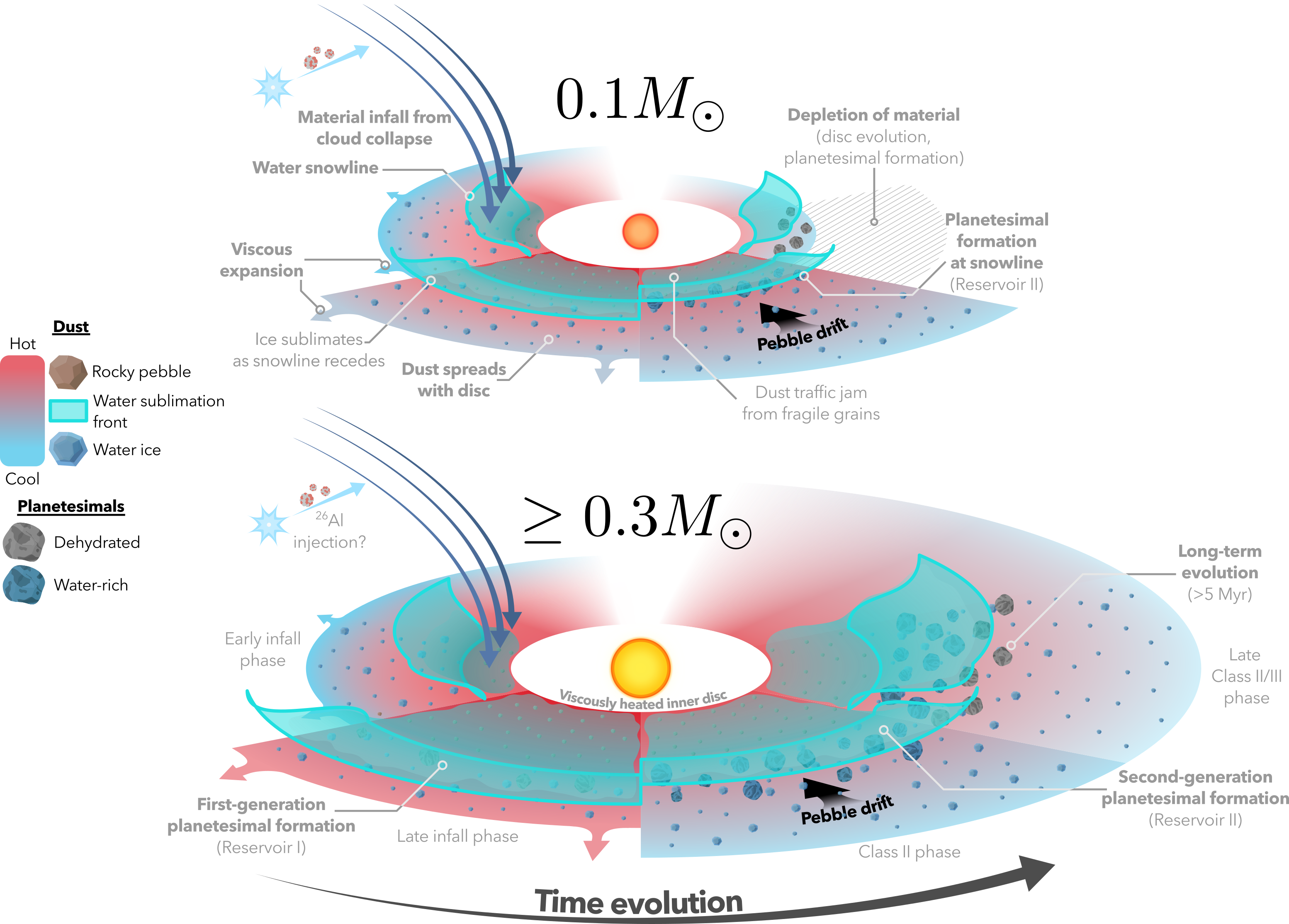}
    \caption{Schematic illustration of the different evolutionary stages for discs born from $M_{\rm{cloud}}=0.1 M_\odot$ and $M_{\rm{cloud}}\geq 0.3 M_\odot$ clouds. \textit{Top:} evolution of a forming disc around a low-mass M-dwarf. $^{26}$Al-injected material infalls from the collapsing cloud and feeds the disc for a brief period, pushing the snowline outward due to viscous heating. The disc viscously expands, dust moving with it, before dust coagulates into icy pebbles. These pebbles drift and deliver their ices to the water snowline, where planetesimals form. Beyond a few Myr, the dust disc disappears due to dynamical evolution, and planetesimals are dehydrated by $^{26}$Al. \textit{Bottom:} as above, but for stars born from $M_{\rm{cloud}}\geq 0.3 M_\odot$ clouds. Discs around these stars form planetesimals during the infall phase, possibly akin to the proto-Sun \citep{lichtenberg2021_bifurcation}, before pebble drift leads to more planetesimal formation at the snowline. The disc then continues to evolve in the Class II/III phase.}
    \label{fig:infall schematic}
\end{figure*}

There is a clear relationship between cloud mass and rotation rate, stellar mass, and timings and outcomes of planetesimal formation. Discs forming from $0.1M_\odot$ mass clouds evolve significantly faster and produce planetesimals in a distinctly different way to higher mass discs; we summarise the contrast in formation pathways in Fig.~\ref{fig:infall schematic}. In the following sections, we discuss our results and their implications for planet formation and disc evolution, as well as consider additional cloud parameters.

\subsection{Variations in Planetesimal Composition}
\label{subsubsec:planetesimal reservoirs}

\begin{figure}
	\includegraphics[width=1\columnwidth]{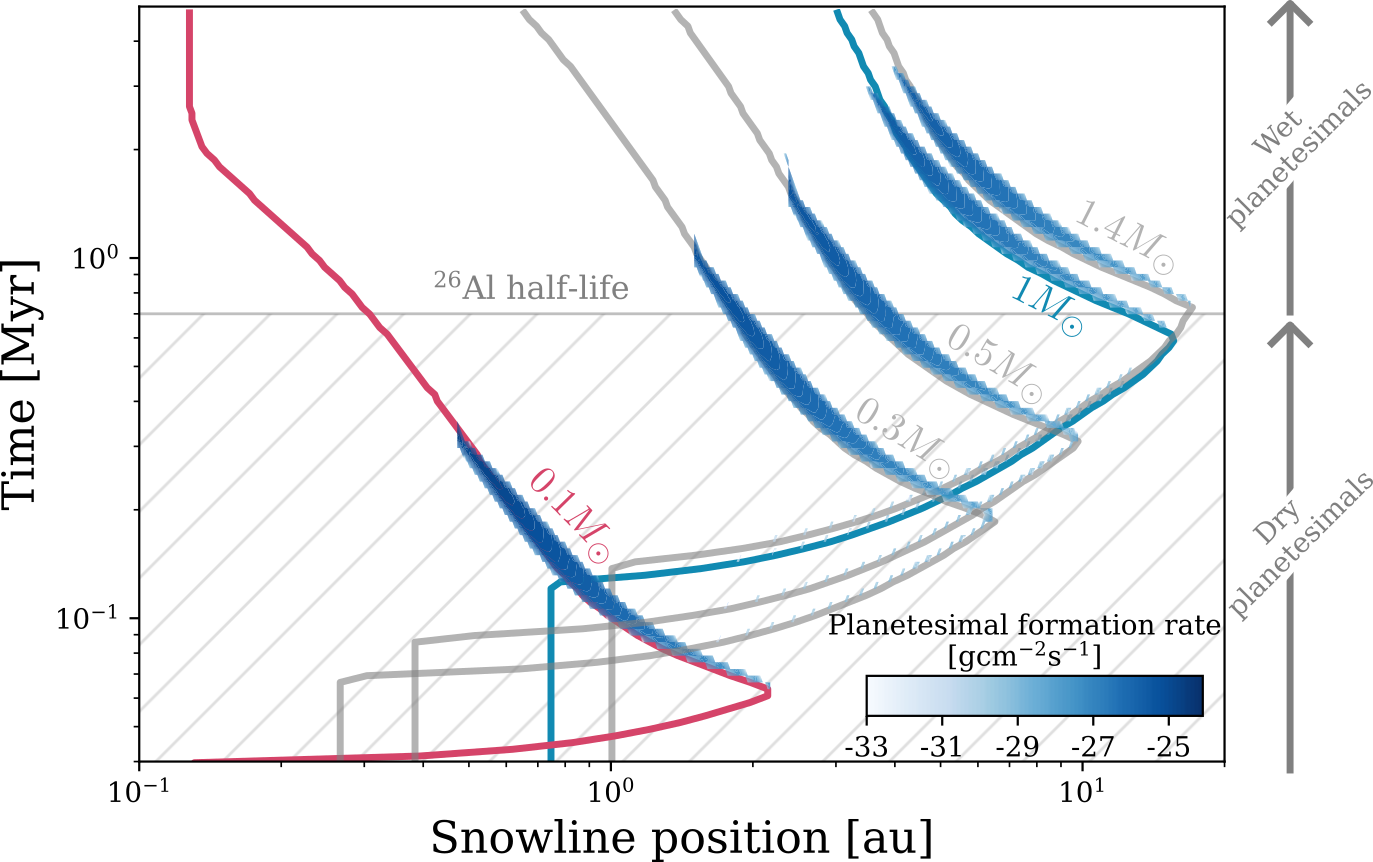}
    \caption{Planetesimal formation rate and snowline position as a function of radial distance and time, for different cloud masses and radii. The half life of $^{26}$Al, $\tau_{\rm{^{26}Al}}$, a radioactive isotope that may be responsible for significant planetesimal heating in the protosolar nebula, is shown with the hatched region (see text for details). The disc in the $0.1M_\odot$ system produces its planetesimals exclusively in $t<\tau_{^{26}\rm{Al}}$, meaning that planetesimals that will be dehydrated.}
    \label{fig:Al26 timings}
\end{figure}

Planetesimal formation in protoplanetary discs is sensitive to the formation timing of the host disc and star. Reservoir I can form very early in the disc formation period, the timing varying with stellar mass, whereas later-formed planetesimals emerge in the the Class II phase, the timing also varying with stellar mass. When exactly planetesimals form relative to CAI formation has important ramifications on planetesimal volatile compositions due to radiogenic heating. Analysis of the meteoritic oxidation states indicates the inclusion of water-rich content into NC-like planetesimals \citep{grewal2024_water_rich_NCs}, suggesting that early-forming planetesimals were indeed created in the inner disc. In this and the following section, we focus on $^{26}$Al as this is the driver for planetesimal heating for Solar System-like abundances \citep{nittler&ciesla2016_extraterrestrial, lugaro2018_al26_variability, parker2020_planet_birth}, which may be representative of the galactic mean \citep{jura&young2014_extrasolar, fujimoto2018_galactic_isotopes}; however, it may be uncommon for solar-like nebulae to be injected with $^{26}$Al \citep{parker2014_al26_solar_enrichment, eatson2024_al26_injection}.

Planetesimals birthed within the half life of $^{26}$Al  ($t<\tau_{^{26}\rm{Al}}$, $\sim700~\rm{kyr}$, L21) will experience radiogenic heating, whereas planetesimals created later ($t\gtrsim \tau_{^{26}\rm{Al}}$) will not; the heated planetesimals will become dehydrated, although smaller planetesimals will instead cool (L21). This creates a chemical heterogeneity in the system, which has been argued to be partly responsible for the CC/NC dichotomy in the Solar System (L21). 

This chemical heterogeneity is not unique to the solar system based on our models. Every simulation forms planetesimals in the infall and Class II phase, \textit{except} for the $0.1M_\odot$ cloud simulation, which only forms a single reservoir in its Class II phase. This Class II phase, however, occurs in $t<\tau_{^{26}\rm{Al}}$ due to the short-lived infall phase followed by rapid disc evolution. We illustrate this in Fig.~\ref{fig:Al26 timings}, where we show the snowline position and planetesimal formation in the context of $^{26}$Al. This result is robust against variable cloud rotation rates (see Section~\ref{subsubsec:planetesimal formation variation}).

This rapid planetesimal formation around the $0.1M_\odot$ star means all of the planetesimals will be chemically \textit{homogeneous} instead of heterogeneous, as around higher-mass stars. These planetesimals around low-mass M-dwarfs ($M_\star=0.1M_\odot$) will all devolatise and become dehydrated from the subsequent $^{26}$Al heating, if present. This has consequences for the volatile content of emergent planets around very low mass stars. We discuss this in the following section.

\subsection{Do Planets Form Dry Around M-dwarf Stars?}
\label{subsec:mdwarf planets}

The timing of planetesimal formation around low-mass M-dwarf stars ($M_\star=0.1M_\odot$) has important implications for the volatile content of the subsequently forming planets.

In our simulations, the material available for planet formation around M-dwarf stars is rapidly locked up into planetesimals and the remaining dust and ice is quickly depleted through pebble drift, all in $<2~\rm{Myr}$. The rapid formation of planetesimals occurs within 500 kyr, less than the half-life of $^{26}$Al ($\sim700~\rm{kyr}$, L21), meaning that all planetesimals will contain radioactively decaying $^{26}$Al, if it is injected into the protostellar nebula. The internal evolution of these planetesimals will be similar to Reservoir I planetesimals formed in the Solar System (L21) - they will be dry due to internal radiogenic heating. 

Despite their early formation, the planetesimals may initially lack sufficient mass to accrete any pebbles without mutual collisional growth \citep{visser&ormel2016_planetesimal_growth, liu2019_planetesimal_to_pebble_accretion}. The pebble flux should be $\dot{M}\geq 10^{-6}~\rm{M_{\oplus}}~\rm{yr}^{-1}$ to be high enough for efficient growth by pebble accretion \citep[][L21]{levison2015_terrestrial_growth}. During this brief phase, there is a total of 0.18$M_{\oplus}$ of material available for accretion.

The amount of pebbles accreted depends on the pebble accretion efficiency \citep{ormel&liu2018}. Following L21, we suppose an embryo mass of $3.11\times 10^{23}~\rm{g}\approx 5.2\times10^{-5}~\rm{M_{\oplus}}$, giving planet-to-star mass ratio of $q=1.6\times10^{-9}$, and disc aspect ratio of $h\sim0.1$, typical up to $\sim$10 au for our M-dwarf disc. For a typical pebble Stokes number of $\rm{St}\sim10^{-3}$, $\alpha_{\rm{z}}=\alpha_{\rm{turb}}=10^{-5}$, and gas pressure gradient of $10^{-3}$, the pebble accretion efficiency of such an embryo is much less than a percent \citep[$\sim10^{-5}$ based on][]{ormel&liu2018}. This means collisional growth is required to reach pebble accretion-efficient masses. However, the planetesimals may not grow quickly enough via mutual collisions to make use of the drifting pebbles before they disappear by $\sim 2~\rm{Myr}$, meaning emerging planet compositions will be dictated by the content of planetesimals. Therefore, any rocky planets that form around low-mass M-dwarf stars will inherit the dehydrated compositions and will be volatile-poor and later lack a significant atmosphere.

This prediction is in agreement with the lack of rocky planet atmospheres observed around M-dwarfs with JWST \citep[e.g.][]{kreidberg&stevenson2025_JWST_rocky_atmo} and independent models pointing to volatile-poor formation of e.g. TRAPPIST-1c \citep{teixeira2024_trappist1c_formation}. Further constraints on the balance between volatile escape and delivery from planetary interiors may be required to fully understand the volatile content of planets around M-dwarfs \citep{lichtenberg2025_exo_interior_obs}.

The precise chemical composition of these planetesimals will depend strongly on the material content that is inherited from the parent cloud, and the size of planetesimals (L21). The inherited material is dependent on the assumptions in protoplanetary disc models about ice composition and processing \citep[e.g.][]{schneider&bitsch2021a_chemcomp, mah2023_vlms_C/O, houge2025_organics}, chemical evolution \citep[e.g.][]{booth&ilee2019_pebble_chem, eistrup2016_volatile_evolution, eistrup2018_evolving_C/O, sellek&van_dishoeck2025_carbon_chemistry} and compositional mixing \citep[e.g.][]{bergner2024_jwst_ice, williams2025_entrapment}. The amount of internal radiogenic heating \citep{alexander2018_meteorite_water, lichtenberg2019_water_dichotomy} and planetesimal size  \citep[][L21]{lichtenberg2018_planetesimal_evolution} will additionally impact planet and planetesimal composition; the former of which is dependent on $^{26}$Al abundance, which may vary greatly across exoplanetary systems and star-forming zones \citep{lugaro2018_al26_variability}. The exact nature of the evolution of these potentially dry planetesimals around M-dwarf stars is beyond the scope of this work and will therefore be the subject of a future study.

\subsection{Sensitivity to Cloud and Disc Parameters}
\label{subsec:variable cloud parameters}

We have thus far investigated the combined effect of cloud mass and stellar luminosity, and we now look to rescaling the cloud rotation rate to expand the parameter space. We note that the most massive clouds $M>1M_\odot$ with the highest rotation rates can create gravitationally unstable discs due to the disc-to-star mass ratio \citep{kratter&lodato2016} and therefore may not produce planetesimals in the way our models predict, and encourage readers to analyse the results for $\Omega_{\rm{cloud}}=4\Omega_{\rm{DD18}}$ and $M_{\rm{cloud}}\geq1M_\odot$ with caution. We discuss this further in Section~\ref{subsubsec:GI limitation}.

\subsubsection{Planetesimal Formation}
\label{subsubsec:planetesimal formation variation}

\begin{figure}
	\includegraphics[width=1\columnwidth]{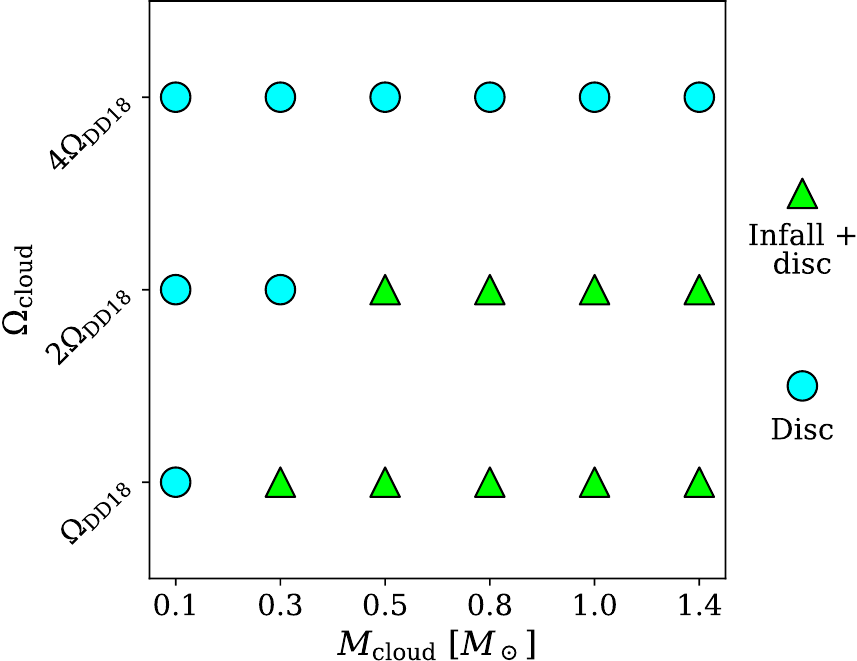}
    \caption{When planetesimals are formed (during infall and in the disc phase, or just the disc phase) depending on cloud mass and rotation rate. Generally, systems that produce one reservoir of planetesimals do so in the Class II disc phase, and multi-reservoir systems produce them in both infall and the Class II phase.}
    \label{fig:parameter space}
\end{figure}

All scenarios we investigate form planetesimals in either both the infall and Class II disc phase, or just the Class II disc phase (see Fig.~\ref{fig:parameter space}). All systems except for the $0.1M_\odot$ cloud produce planetesimals during the infall phase for $\Omega_{\rm{cloud}} = \Omega_{\rm{DD18}}$, meaning all but this system form Reservoir I. As the cloud rotation rate doubles, clouds with masses up to $0.5M_\odot$ do not produce planetesimals during the infall phase as the infall occurs earlier and the snowline migrates further out, whereas more massive clouds still meet the streaming instability criteria during infall. When the cloud rotation rate is increased further, however, no planetesimals are formed during the infall phase in any system. We show the variation in snowline position in Figure~\ref{fig:snowline variation} in Appendix~\ref{appendix:snowline variation}.

We also note that the total planetesimal mass systematically increases with increasing cloud rotation rate, which we show in Fig.~\ref{fig:planetesimal mass} in Appendix~\ref{appendix:mass reservoirs}. This is likely because the centrifugal radius of clouds with higher rotation rates is larger, meaning solid material is deposited further from the star. This gives the snowline access to a larger reservoir of dust and ice, both of which are important for planetesimal formation during the Class II phase \citep[][DD18]{drazkowska&alibert2017_snowline_plts}.

Clouds with lower angular momentum compared to observations ($\Omega_{\rm{cloud}}=\Omega_{\rm{DD18}}$ in Fig.~\ref{fig:cloud initial conditions}) are more likely to produce planetesimals during molecular cloud collapse, whereas higher rotation rates impede early planetesimal formation. Despite this, clouds with higher rotation rates manage to  produce more planetesimals (see rightmost panel of Fig~\ref{fig:planetesimal mass}). This suggests that the Solar System is likely to have formed from a lower angular momentum cloud ($\Omega_{\rm{cloud}}\leq 2\Omega_{\rm{DD18}}$). In addition, planetesimals form at approximately the same time and in the same disc evolutionary stage between cloud rotation rates; in fact, planetesimals are always formed within the $^{26}$Al half-life in the disc emerging from the $0.1M_\odot$ cloud. Our discussion about dehydrated planet(esimal) formation is therefore robust against different molecular cloud parameters.

\subsubsection{Planetesimal Belt Widths}

\begin{figure}
	\includegraphics[width=1\columnwidth]{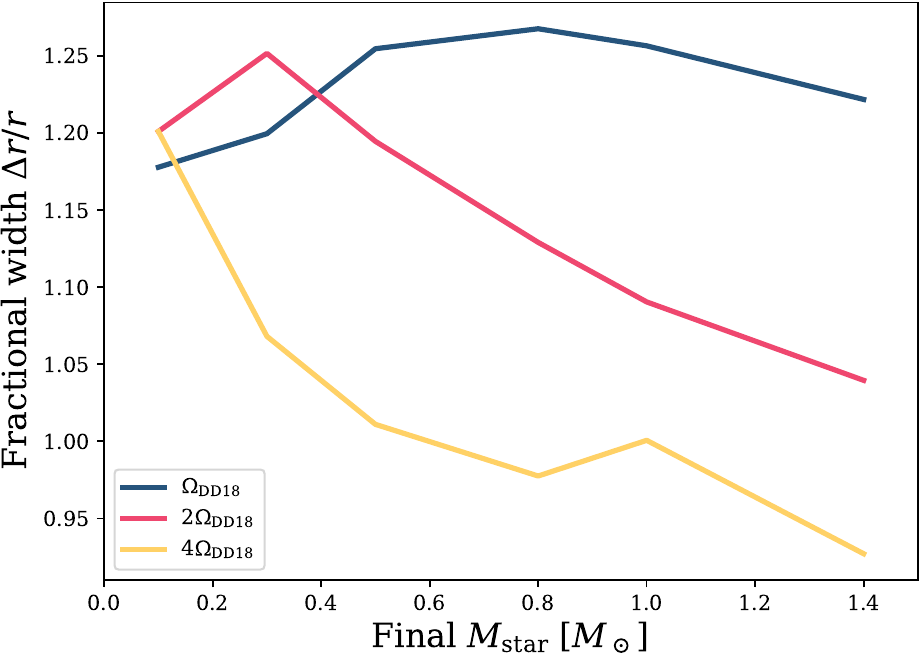}
    \caption{The fractional width of the planetesimal surface density as a function of cloud mass. The different lines show the effects of different cloud rotation rates (see Fig.~\ref{fig:cloud initial conditions}). Higher rotation rates correspond to radially larger clouds and discs, generally decreasing the fractional width of the planetesimal distribution.}
    \label{fig:planetesimal belts}
\end{figure}

We now briefly discuss the impact of cloud mass and rotation rate on planetesimal belt fractional width, i.e. the radial width of the belt divided by the distance from the belt to the star. We define the belt width $\Delta r$ as the width of the planetesimal mass distribution $\Sigma_{\rm{plts}}$ that contains 95\% of the total planetesimal mass, and the distance $r$ to be the median of the planetesimal distribution (such that 50\% of the planetesimal mass lies on either side of the median).

We find belt fractional widths consistent with the upper end of fractional width distributions from recent observational surveys of debris discs (REASONS, \citealt{matra2025_REASONS}, and ARKS, \citealt{marino2026_ARKS_overview, han2026_ARKS_radial_structure}). We show the fractional widths from our simulations in Fig.~\ref{fig:planetesimal belts}. For $\Omega_{\rm{DD18}}$, as the cloud mass and rotation rate increase, the fractional width increases up to $0.8M_\odot$ before decreasing again. We attribute the increasing width with increased centrifugal radius and the water snowline migrating at a different rate. For clouds with mass $M>0.8M_\odot$, the snowline migrates further out, lowering the belt fractional width.

Clouds with higher rotation rates see a decreasing belt width with final stellar mass. This is likely due to the fact that fewer discs produce planetesimals in the infall phase, and the snowline migrates to a similar position for $M_\star>0.8M_\odot$ but migrates at different rates, thereby producing belts of different widths. We note, however, that these high rotation rates are unlikely to produce gravitationally stable discs and therefore be observable as smooth, planetesimal-building discs (see Section~\ref{subsubsec:GI limitation} for further discussion).

The planetesimals belts in our simulations have fractional width $\sim1$, which is consistent with the upper limits of observed fractional widths, and is higher than many other observed belts \citep[e.g.][]{han2026_ARKS_radial_structure}. This indicates that dynamical interactions between planetesimals and planets may be required to sculpt their shape, if the observed debris discs are leftover from planetesimal formation at the migrating snowline \citep[e.g.][]{kennedy&wyatt2010_dynamical_discs, pearce2024_planet_sculpting, bendahan-west2026_planet_search}. This dynamical sculpting will likely erase the belt's memory of the disc's thermal history. Furthermore, observed debris discs may not have been formed at the snowline, but instead processes like external photoevaporation (Johnston et al. in prep).

\subsection{Disc Evolution and Setting $t=0$}

Traditional disc evolution models of Class II protoplanetary discs typically select $t=0$ as the beginning of dust evolution in a pre-formed disc \citep[e.g.][]{birnstiel2012_twopoppy, drazkowska2021, williams2025_entrapment}. We, however, select the start of the molecular cloud collapse as $t=0$ (see Section~\ref{subsec:disc model}). Here, we discuss what the choice of $t=0$ means for discs around different stellar types in the context of the $\dot{M}_{\rm{acc}}-M_\star$ relation and dust evolution.

\subsubsection{Reproducing the Observed $\dot{M}_{\rm{acc}}-M_{\star}$ Relation}
\label{subsec:mdot-m relation}

\begin{figure}
	\includegraphics[width=1\columnwidth]{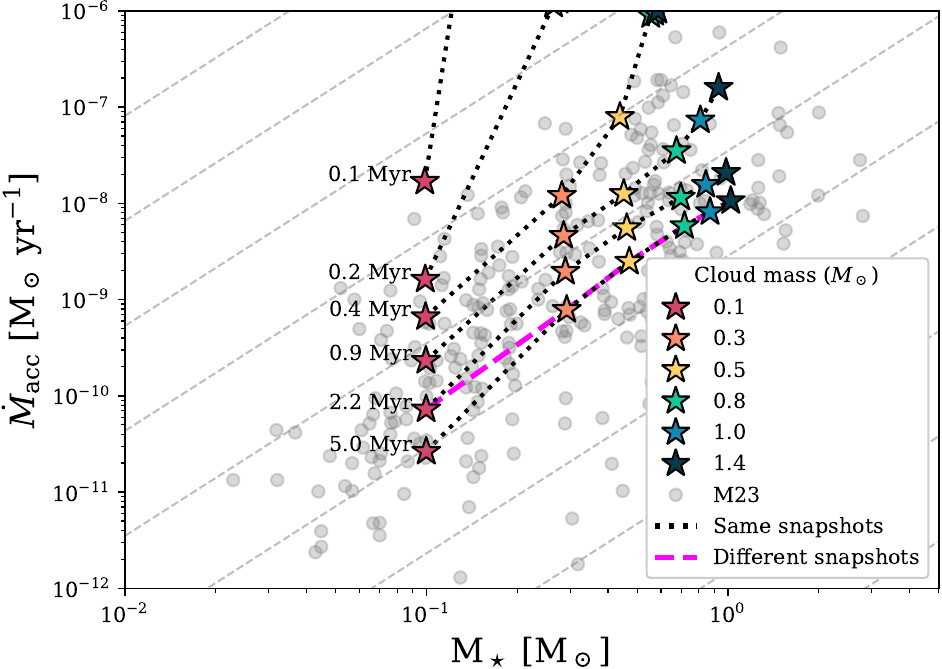}
    \caption{Observed (grey circles) and simulated (coloured stars) steeper-than-linear relationship between $\dot{M}_{\rm{acc}}-M_\star$ with grey dashed lines showing $\dot{M}_{\rm{acc}}\propto M_\star^2$ for visual aid. These lines do not represent a fit to the data, which has been adapted from Fig.~4 of \citet{manara2023_pp7_young_stars} and references therein. We connect the same time snapshots between simulations with black dotted lines to create temporal isochrones, and connect them by a magenta dashed line to show an isochrone of discs at different time snapshots that might recover the observed trend. Each system is in the Class II phase by 2.2 Myr.}
    \label{fig:mdot-m relation}
\end{figure}

We compare our simulations to observational trends in Fig.~\ref{fig:mdot-m relation}, where we plot the stellar mass accretion rate $\dot{M}_\star$ against stellar mass $M_\star$ for each simulation. We also plot data adapted from Fig.~4 of \citet{manara2023_pp7_young_stars} and references therein for comparison. We additionally link the same temporal snapshot in each simulation with black dotted lines to create isochrones, which can be used to compare against $\dot{M}_{\rm{acc}} \propto M_\star^{2}$, although we note that this relation is not a fit to the observational data \citep{manara2023_pp7_young_stars}.

Our models match the observed parameter space well. When connecting each model at the same time $t$ in the simulation via an isochrone (labelled `Time snapshot' in Fig.~\ref{fig:mdot-m relation}), the models follow a steeper-than-linear relation, as observed in the literature \citep[e.g.][]{hillenbrand1992_herbig_discs, muzerolle2003_yso_accretion, mohanty2005_t_tauri, natta2006_ophiucus_accretion, manara2023_pp7_young_stars}. This relation matches closely to $\dot{M}_{\rm{acc}} \propto M_\star^{2}$ for $M_{\star}\geq0.3M_\odot$. This relationship, however, breaks down when including low mass stars ($0.1M_\odot$). This suggests that discs around low mass stars may have accreted more of their initial disc material and hence be at a later evolutionary stage, despite the same amount of time having elapsed since our chosen $t=0$.

Each disc evolves at different rates: the $0.1M_\odot$ system rapidly accretes its dust disc after forming planetesimals due to rapid evolution around low mass stars \citep[][also see Section~\ref{subsubsec:dust reservoirs}]{pinilla2022_vlms_drift}. Therefore, the $0.1M_\odot$ system is at a much later evolutionary stage than the $1M_\odot$ system, for instance, despite being the same age $t$. We illustrate that the constant gradient of the steeper-than-linear relation given by \citet{manara2023_pp7_young_stars} can be recovered if the low-mass star is connected at an arbitrarily chosen earlier (i.e. younger) time, which may represent an earlier stage of the disc's evolution and its lifetime when accretion rates are higher. This is shown by the magenta dashed line in Fig.~\ref{fig:mdot-m relation}, marked as `different snapshots'. This indicates that discs that have evolved for the same length of time since the chosen $t=0$ (i.e. same snapshot) may have reached different stages of their lifetime. This effect is particularly strong for low-mass stars, as their discs evolve on much shorter timescales. This evolution may be made manifest on observables used to determine protoplanetary disc evolutionary stage, such as SEDs, although understanding how the differing disc evolution impacts SEDs requires further detailed modelling beyond the scope of this work.

We therefore conclude that our simulations fit well to observational trends when connecting their evolutionary stages, both a priori through our initial conditions (Fig.~\ref{fig:cloud initial conditions}) and a posteriori through the $\dot{M}_{\rm{acc}} - M_\star$ relation. They highlight, however, that discs at the same time after their birth may not be at the same evolutionary stage.

\subsubsection{History of Dust Evolution}
\label{subsec:pebble drift}

\begin{figure*}
	\includegraphics[width=1\textwidth]{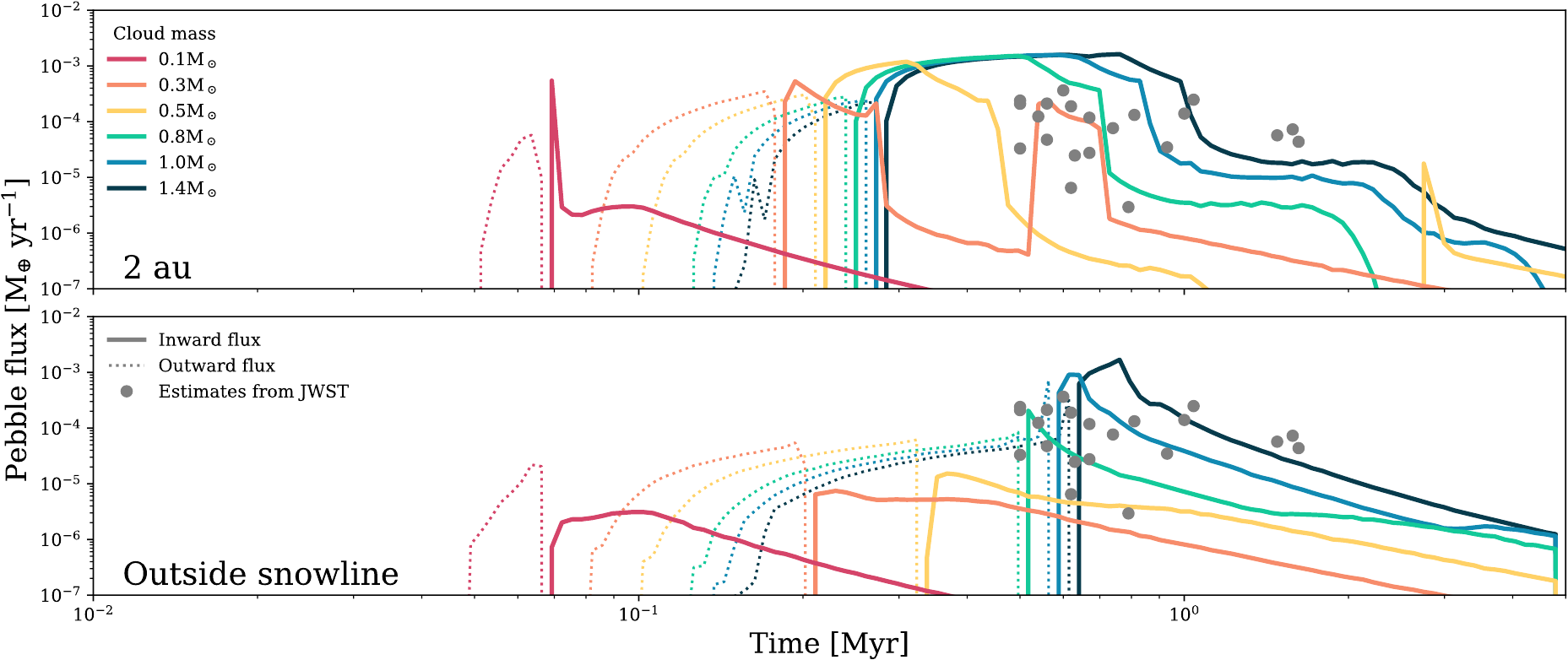}
    \caption{Pebble flux through 2 au (top) and just outside the snowline (bottom), the latter of which is a function of time, stellar mass, and cloud mass and rotation rate. Each colour denotes a different cloud mass, and the solid lines represent the pebble mass flux travelling radially inwards (i.e. towards the star) whilst the dotted lines are the outward flux (due to diffusion and viscous spreading of the disc). We also show pebble flux estimates from JWST data from \citet{krijt2025_cascades} and references therein for context, but we do not necessarily aim to fit our simulations to this data. The large increase in the pebble flux through 2 au for the $0.3M_\odot$ cloud at 0.5 Myr is due to the snowline migrating through 2 au.}
    \label{fig:pebble flux}
\end{figure*}

Analysing the dust evolution is also a useful way to constrain how discs evolve. Pebble drift and ensuing radial pebble mass are a complex function of space, time, and environment including disc mass and radius \citep{drazkowska2021, williams&krijt2025_mass_constraint}, disc substructure \citep[e.g.][]{kalyaan2021, stammler2023, krijt2025_cascades}, and grain properties \citep[e.g.][]{pinilla2017_icelines, houge2025_smuggled_water, williams&krijt2025_mass_constraint}. Many pebble drift models have historically began with a pre-formed class II disc by using the self-similar surface density profile given by \citet{lynden-bell&pringle1974}, and then calculated the subsequent pebble drift history. When the initial conditions are no longer tightly constrained, the evolution of pebble drift can look quite different across different systems.

We show the pebble flux through the discs in our models as a function of time and position in Fig.~\ref{fig:pebble flux}, 
with estimates of pebble flux inferred from JWST observations of discs around T Tauri stars \citep{krijt2025_cascades} also shown for additional context. The top panel shows the pebble flux through 2 au in each model, whilst the bottom shows the flux through a position just outside the snowline (which varies with time due to accretion heating), showing the material that is delivered to the snowline; this material bears the imprint of the collapsing cloud and the disc's thermal history, telling the story of dust migration. It also dictates the formation of planets, since this material is fed into planetesimals at the snowline \citep[][DD18, L21]{drazkowska&alibert2017_snowline_plts} and possibly later pebble-accreting planets (L21).

We also show the radially outward flux due to the viscous expansion of the disc and dust diffusion in dotted lines. Whilst dust is beginning to grow and drift, the dust migration is dominated by the disc expansion. After a brief expansion period, pebble drift dominates the radial motion of dust.

It is clear that each system evolves on very different timescales: pebble drift begin to dominate dust migration much earlier in the $0.1M_\odot$ case ($\sim0.07$ Myr) and is significantly shorter-lived. In contrast, the more massive systems see pebble drift starting and ending later ($\sim0.6$ and $>5$ Myr for the 1.4$M_\odot$ case). This difference in evolutionary timescales can be explained twofold: firstly, there is more material and a longer infall phase in the more massive clouds. Secondly, the infalling material falls in much closer to the star in the less massive clouds. The ratio of the centrifugal radii of two clouds, $R_{\rm{centr,1}}$ and $R_{\rm{centr,2}}$, where the material is deposited, can be expressed as:

\begin{equation}
    \dfrac{R_{\rm{centr,1}}}{R_{\rm{centr,2}}}=\left( \dfrac{\Omega_{\rm{cloud,1}}}{\Omega_{\rm{cloud,2}}} \right)^{2} \left( \dfrac{M_{\rm{cloud,1}}}{M_{\rm{cloud,2}}} \right)^{3}.
\end{equation}

\noindent The $0.1M_{\rm{cloud}}$ system therefore has a centrifugal radius that is initially 0.01\% of that for the $1M_{\rm{cloud}}$ system. Since viscous, growth, and drift timescales are significantly faster closer to the central star \citep{armitage2013_astrophysics}, the $0.1M_\odot$ system evolves considerably faster. In addition, discs with smaller radii do not experience pebble drift as long as larger discs \citep{williams&krijt2025_mass_constraint}. As a result, the pebble drift phase comes to a halt much more rapidly in the lower mass systems. In general, our results demonstrate the strong impact of modelling both disc formation and dust evolution, and how varying stellar masses can considerably change the evolutionary picture.

\subsection{Model Extensions}
\label{subsec:limitations}

Our simulations demonstrate a clear transition from a single to double reservoir of planetesimals formed at the migrating water snowline across the stellar mass spectrum (see Fig.~\ref{fig:parameter space}). We now discuss additional processes that were not expressly incorporated into our simulations.

\subsubsection{Gravitationally Unstable Discs}
\label{subsubsec:GI limitation}
When the cloud rotation rate is increased dramatically (to $4\Omega_{\rm{DD18}}$), the protoplanetary disc accretes a very significant amount of gas. In fact, the $1M_\odot$ cloud briefly ends up with a disc that is more massive than the protostar - this disc would almost certainly be unstable to the gravitational instability (GI) mechanism \citep{kratter&lodato2016} and would be extremely likely to fragment. GI was not captured in the simulations we ran and is best modelled by full hydrodynamical simulations \citep[e.g.][]{leedham2025_GI_irradiation}. In fact, full 3D models of protoplanetary discs have shown that GI can be induced by infalling material \citep{longarini2025_GI_infall}, indicating the necessity to study the nature of infall using more complex codes than the one presented here. Our 1$M_\odot$ simulation, however, produces a similar final mass of planetesimals to calculations by \citet{huhn2025_3dInfall_plts}, who combine 3D cloud collapse simulations with 1D planetesimal formation, although our formation timescales differ. We therefore emphasise that 1D vertically integrated codes retain their value in studying the effect of infall on certain processes \citep[e.g. planetesimal formation, DD18,][]{zhao2025_plts_by_infall} despite not capturing the full hydrodynamics.

\subsubsection{Impact of Stellar Evolution}
\label{subsubsec:stellar evolution}
We do not evolve the stellar luminosity in our simulations. Although the stellar luminosity can change by up to a factor of 2 for stars of $M_\star<1.4M_\odot$, this would primarily change the position of the water snowline after the accretion heating has dissipated (see the snowline position for the 0.1$M_\odot$ case in Fig.~\ref{fig:snowline position}). This is, however, long after all of the planetesimals have formed: in all simulations, the majority of the planetesimals are formed in the first few hundred thousand years after the cloud has been accreted as the snowline migrates inwards. For the higher mass stars and clouds, the heating from cloud accretion takes longer to dissipate, and so the snowline remains in the accretion heating-dominated regime. As a result, whilst changing stellar luminosity may change where the snowline is, it is unlikely to cause further planetesimals to form. Stellar evolution should therefore not have a significant effect on the results of our simulations.

We picked effective temperatures by comparing the stellar mass of simulations from \citet{baraffe2015_luminosities} to the cloud mass of our simulations, although not all of the cloud mass ends up in the stellar core. Therefore, our snowlines can be expected to be closer into the star by a few au for these higher mass clouds than what we show here. We do not, however, expect that this would impact the main results of our simulations concerning the mass and timing of forming planetesimals, as this is primarily dependent on the cloud mass and angular momentum.

\subsubsection{The Role of Substructure and Planet Formation}
\label{subsubsec:planet formation}

Substructures due to planet formation or other effects may not be present in still-forming Class 0 discs \citep{hsieh2025_CAMPOS_II_substructures}, with a mixture of observations and non-detections in the Class I phase \citep{segura-cox2020, ohashi2023, shoshi_ClassI_gap_cavity}. Class 0 substructures - corresponding to the early infall stage of our models - are likely to be negligible, and the impact of a gap in the Class I phase on planetesimal formation is also likely to be minimal \citep{nazari2025_blanket}. This is because any planet-induced substructure will require planetesimals beforehand to create a planetary embryo. The Reservoir I planetesimals, where formed, are therefore unlikely to be affected.

We note that planetesimal formation may proceed via pathways not included in our simulations \citep[see review by][Sect.~3.1]{drazkowska_pp7_2023}. For example, planetesimal formation via the streaming instability occur in other zones of protoplanetary discs (unrelated to the water snowline), such as pressure bumps \citep{lenz2019_planetesimal_synthesis, carrera2021_planetesimals_in_rings, lau2022_core_bumps, tatarelli2026_plts_in_bumps}, other snowlines and sublimation fronts \citep[e.g.][]{izidoro2022_planetesimals_sublimation}, and dead-zones \citep[e.g.][]{lyra2009_dead_zone_planetesimals}. Additionally, planetesimal formation may even occur at various locations without the streaming instability \citep[e.g.][]{okuzumi2012_porous_planetesimals, krijt2016_planetesimals}. The interplay between planetesimal formation at the snowline and also pressure bumps in or outside the snowline is not fully understood, and additional work is required to achieve a comprehensive picture on this \citep[e.g.][]{zhao2025_plts_by_infall}.

The subsequent formation and growth of planets, however, could complicate the picture. Although we form planetesimals in our model, we do not follow planetesimal evolution and interactions, which could lead to the growth of a planetary core. Once a core has been made, it could accrete pebbles and other planetesimals, and possibly open a gap in the disc. In our model, planetesimal formation in the Class II phase relies on pebble drift to the snowline (DD18, L21), which could be impeded by the presence of a gap \citep{pinilla2012, drazkowska2019, kalyaan2021, mah2024_mind_the_gap}. Other studies, however, suggest that some material may leak through a planet-induced gap \citep[][]{stammler2023, houge2026_how_leaky, gurrutxaga2026_CC_planetesimals}, possibly allowing sufficient material to reach the snowline and continue planetesimal formation, which may be observable in the disc's upper surface layers \citep{houge2025_smuggled_water, krijt2025_cascades, sellek2025_co2_drift}. For example, \citet{krijt2025_cascades} estimate typical pebble mass fluxes between $\sim2$ and $\sim300$ $M_\oplus~\rm{Myr}^{-1}$ for $\sim$Myr old T-Tauri stars, although there is considerable uncertainty on their ages and pebble flux estimates. Extending these estimates from JWST data to lower stellar masses and younger discs (< Myr) would therefore help constrain early-stage disc evolution. As such, the formation of Reservoir II planetesimals may or may not be significantly impacted. More involved models of planet formation, dust evolution \citep[e.g.][]{lau2024_sequential_planets} and disc formation are required to study this effect in depth.

\subsubsection{A Note on Temperature Evolution}
\label{subsubsec:temperature limitation}
The viscous evolution of the disc is performed in a 1D vertically-integrated manner, meaning that calculations of the midplane temperature may not accurately mimic full radiative transfer calculations; a more thorough treatment of self-consistently calculating the midplane temperature may require a 2D model that includes viscous heating and could then alter the planetesimal formation rates and dust distributions within the model (L21). More complex models, such as \texttt{cuDisc} \citep{robinson2024_cudisc} or other hydrodynamical codes like \texttt{Phantom} \citep{price2018_phantom}, would be appropriate in performing such calculations.

\section{Conclusions}
\label{sec:conclusions}

We present a new suite of simulations of molecular cloud collapse and protoplanetary disc build-up, including planetesimal formation, which expand upon existing work \citep{drazkowska&dullemond2018_disk_buildup, lichtenberg2021_bifurcation}. We extend these models to sub- and super-solar mass clouds and explore how different cloud parameters influence planetesimal formation. Our main conclusions are as follows:

\begin{enumerate}
\item Planetesimal formation in emerging discs varies in position, timing, and outcomes across the stellar mass spectrum. They form extremely rapidly ($<500$ kyr) around low-mass M-dwarf stars ($0.1M_\odot$), and form in both the infall and Class II phase of discs around stars of mass $M_\star>0.1M_\odot$. More massive stars with more massive discs create more planetesimals.
    \item Planetesimals formed at the water snowline will likely be chemically heterogeneous across the stellar mass spectrum, built from two populations: dry planetesimals heated by $^{26}$Al, and wet planetesimals that form too late. Planetesimals around low-mass M-dwarf stars, however, form planetesimals so rapidly in all our simulations that they are likely to be dry. 
    \item Exoplanets will likely form volatile-poor around low-mass M-dwarfs. Bodies built from early-forming, dehydrated planetesimals will inherit their composition, potentially leading to barren rocky worlds around low-mass M-dwarfs and explaining the lack of atmospheres seen by JWST.
    \item Protoplanetary discs evolve at different times and rates across stellar mass. low-mass M-dwarfs evolve on significantly shorter timescales than their higher mass counterparts: infall, planetesimal formation, and pebble drift occurs more rapidly, depriving planetary embryos of further pebbles and forcing planetary growth to occur via collision-driven mechanisms. More massive stars and discs, however, see planetesimal formation and pebble drift occur later, indicating that finding a common $t=0$ across stellar masses is non-trivial.
    \item The final architectures of planetary systems arising from the planetesimal distributions are keenly sensitive to the initial conditions of the collapsing molecular cloud. More massive clouds exhibit wider planetesimal belts than their smaller counterparts and provide more sustained pebble drift for planetary growth.
\end{enumerate}

Our models have important ramifications for exoplanet formation around low-mass M-dwarf stars ($0.1M_\odot$). Studying the volatile content evolution of the emergent planetesimals foundational to exoplanets will be therefore subject of future work.    

\section*{Acknowledgements}

The authors would like to thank the anonymous reviewer for their helpful comments and suggestions that helped improve the manuscript. JW is funded by the UK Science and Technology Facilities Council (STFC), grant code ST/Y509383/1. JW and SK would like to thank Matthew Bate, Heather Johnston, and Sebastian Marino for helpful suggestions and insightful conversations that helped shape this work. JD was funded by the European Union under the European Union’s Horizon Europe Research \& Innovation Programme 101040037 (PLANETOIDS). Views and opinions expressed are however those of the authors only and do not necessarily reflect those of the European Union or the European Research Council. Neither the European Union nor the granting authority can be held responsible for them. TL was supported by the Branco Weiss Foundation, the Netherlands eScience Center (PROTEUS project, NLESC.OEC.2023.017), the Alfred P. Sloan Foundation (AEThER project, G202114194), NASA’s Nexus for Exoplanet System Science research coordination network (Alien Earths project, 80NSSC21K0593), and the NWO NWA-ORC PRELIFE Consortium (PRELIFE project, NWA.1630.23.013).

\section*{Data Availability}

The data underlying this article will be shared on reasonable request to the corresponding author.

\bibliographystyle{mnras}
\bibliography{bibfile}

\appendix

\section{Derivation of Equation 2}
\label{appendix:omega scaling derivation}

A cloud's specific angular momentum can be expressed as $J=\Omega R^2$. We would like to change the cloud's angular rotation rate $\Omega$ such that the specific angular momentum follows the relation $J\propto R^{1.5}$ \citep[e.g.][]{chen&ostriker2018_cloud_J}. We therefore stipulate:

\begin{equation}\label{eqn:J propto R}
    J=\Omega R^2 \propto R^{1.5}.
\end{equation}

To avoid finding the constant of proportionality, we choose some reference cloud parameters $\Omega_0$ and $R_0$ (see Section~\ref{subsec:cloud initial conditions} for the values we choose) and can therefore write:

\begin{equation}\label{eqn:J propto R new cloud}
    J_0 = \Omega_0 R_0^2 \propto R_0^{1.5}.
\end{equation}

For a new cloud with a new pre-defined mass $M_{\rm{new}}$ and radius $R_{\rm{new}}$, we can find its angular momentum $J_{\rm{new}}$ by dividing equation~\ref{eqn:J propto R} by equation~\ref{eqn:J propto R new cloud}:

\begin{equation}
    \dfrac{J_{\rm{new}}}{J_0}=\dfrac{\Omega_{\rm{new}}R^{2}_{\rm{new}}}{\Omega_0 R^2_0}=\left(\dfrac{R_{\rm{new}}}{R_0}\right)^{1.5}.
\end{equation}

\noindent We can now rearrange this for $\Omega_{\rm{new}}$:

\begin{equation}
    \Omega_{\rm{new}} = \Omega_0 \left( \dfrac{R_{\rm{new}}}{R_0} \right)^{-0.5},
\end{equation}

\noindent which is equation~\ref{eqn:Omega rescaling}.

\section{Comparison to L21}
\label{appendix:comparison to L21}

\begin{figure}
	\includegraphics[width=1\columnwidth]{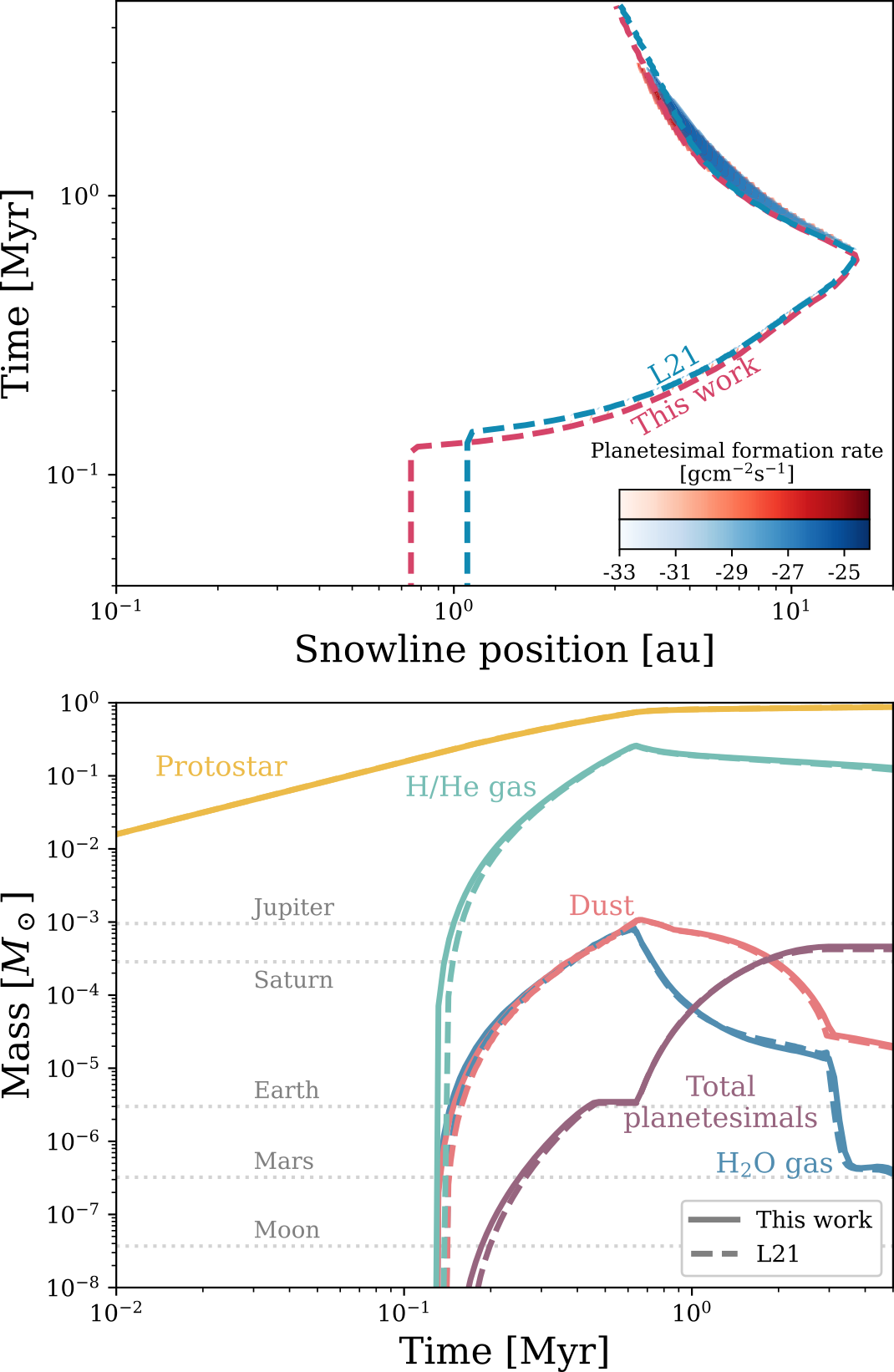}
    \caption{Comparison between our $M_{\rm{cloud}}=M_\odot$ simulation to the updated results for \citet[][labelled L21]{lichtenberg2021_bifurcation}, following corrections from \citet{drazkowska&dullemond_2023_corrigendum}. \textit{Top:} Position of the water snowline as a function of time, following Fig.~\ref{fig:snowline position}. \textit{Bottom:} Mass reservoir evolution, following Fig.~\ref{fig:mass reservoirs}.}
    \label{fig:L21 comparison}
\end{figure}

\section{Dust Evolution ($0.1$, $0.5$, $1M_\odot$)}
\label{appendix:dust maps}

\begin{figure*}
	\includegraphics[width=1\textwidth]{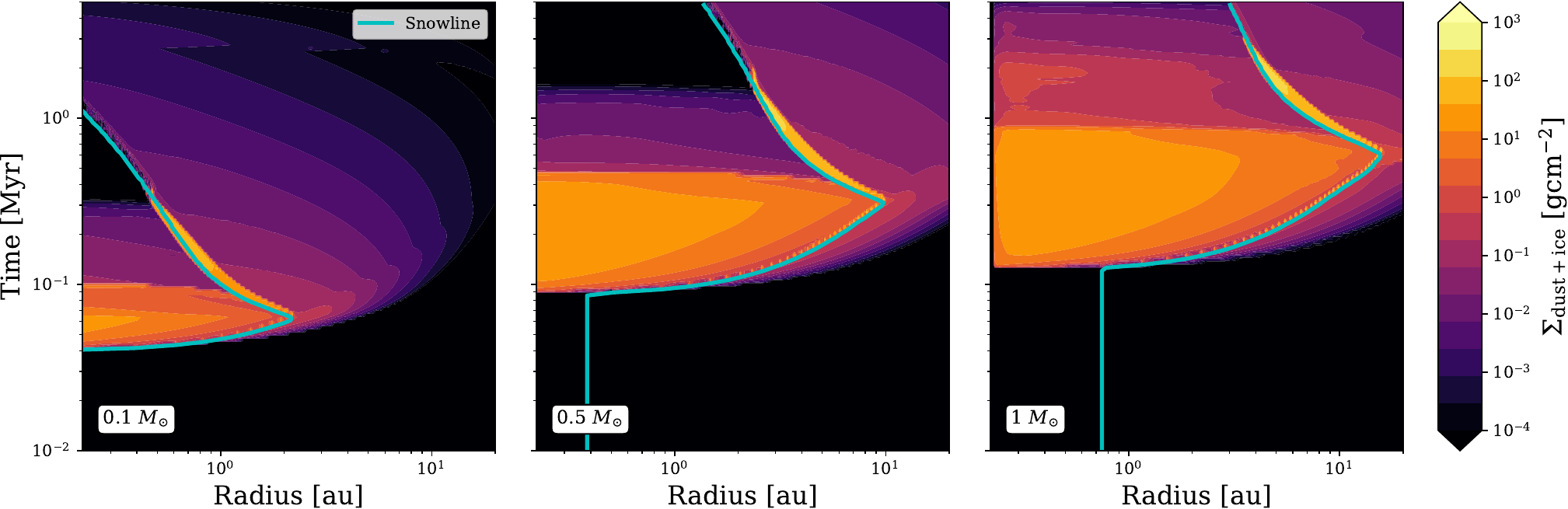}
    \caption{Evolution of the dust and ice surface density as a function of radius and time for the same cloud masses shown in Fig.~\ref{fig:mass reservoirs}. The snowline is overlaid in cyan for reference. The dust overdensities outside the snowline in the Class II phase can be identified as where the streaming instability criteria is met, leading to planetesimal formation.}
    \label{fig:dust maps}
\end{figure*}

\section{Snowlines in Different Clouds}
\label{appendix:snowline variation}

\begin{figure*}
	\includegraphics[width=1\textwidth]{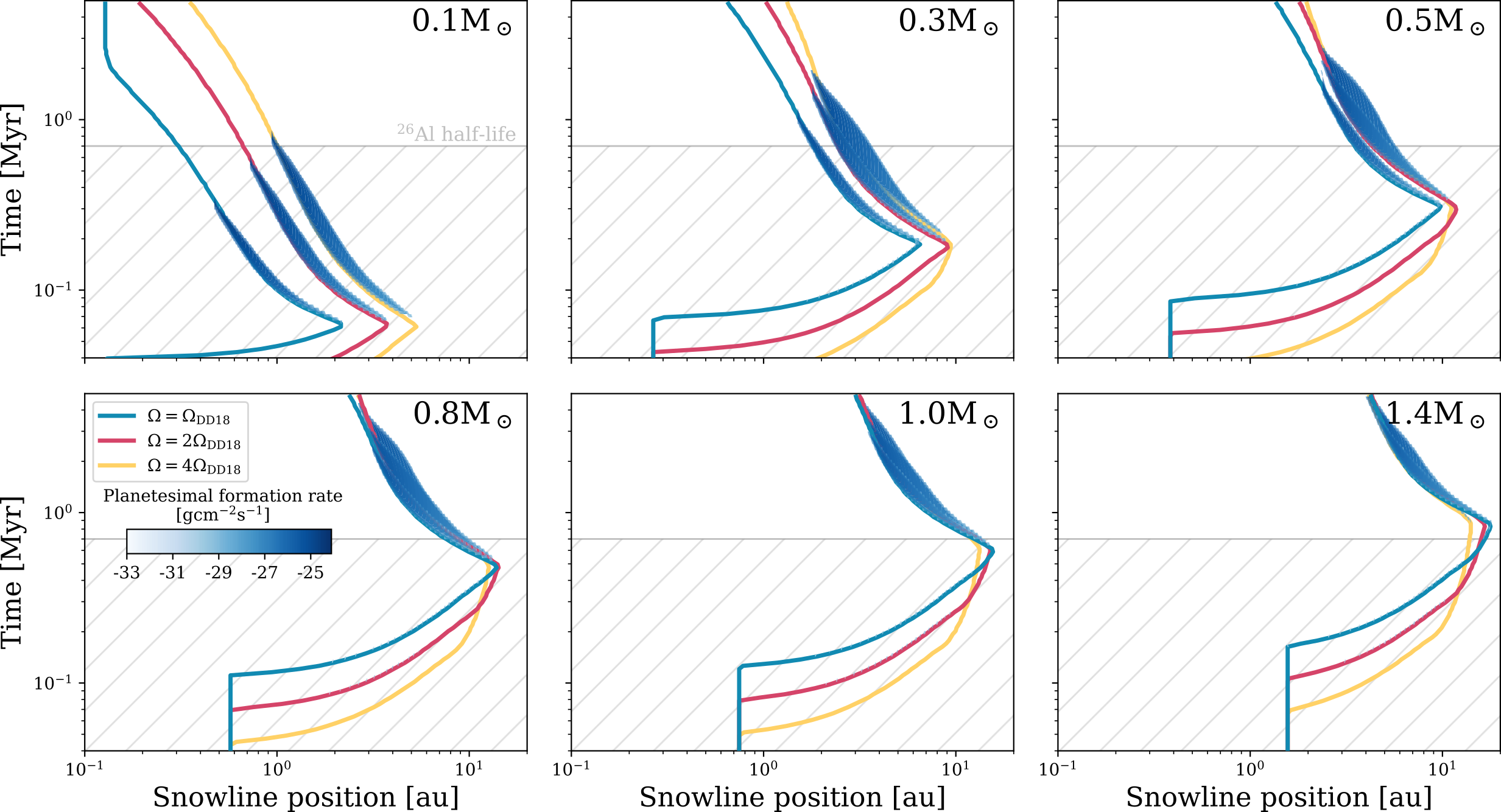}
    \caption{Position of the snowline over time for different cloud masses (panels, labelled) and cloud rotation rates (colours; blue for $\Omega_{\rm{DD18}}$, red for $2\Omega_{\rm{DD18}}$, and yellow for $4\Omega_{\rm{DD18}}$. Higher cloud rotation rates cause infall to occur sooner, the disc is heated up more significantly (pushing the snowline farther out), but generally maintain planetesimal formation in the Class II phase (shown by colour map). The $^{26}$Al half-life is shown in the grey hatched zone. Planetesimal formation always occurs within $t<\tau_{^{26}\rm{Al}}$ for the $0.1M_\odot$ simulation.}
    \label{fig:snowline variation}
\end{figure*}

\section{Individual Mass Reservoirs}
\label{appendix:mass reservoirs}

\subsection{Planetesimal Mass}
\begin{figure*}
	\includegraphics[width=1\textwidth]{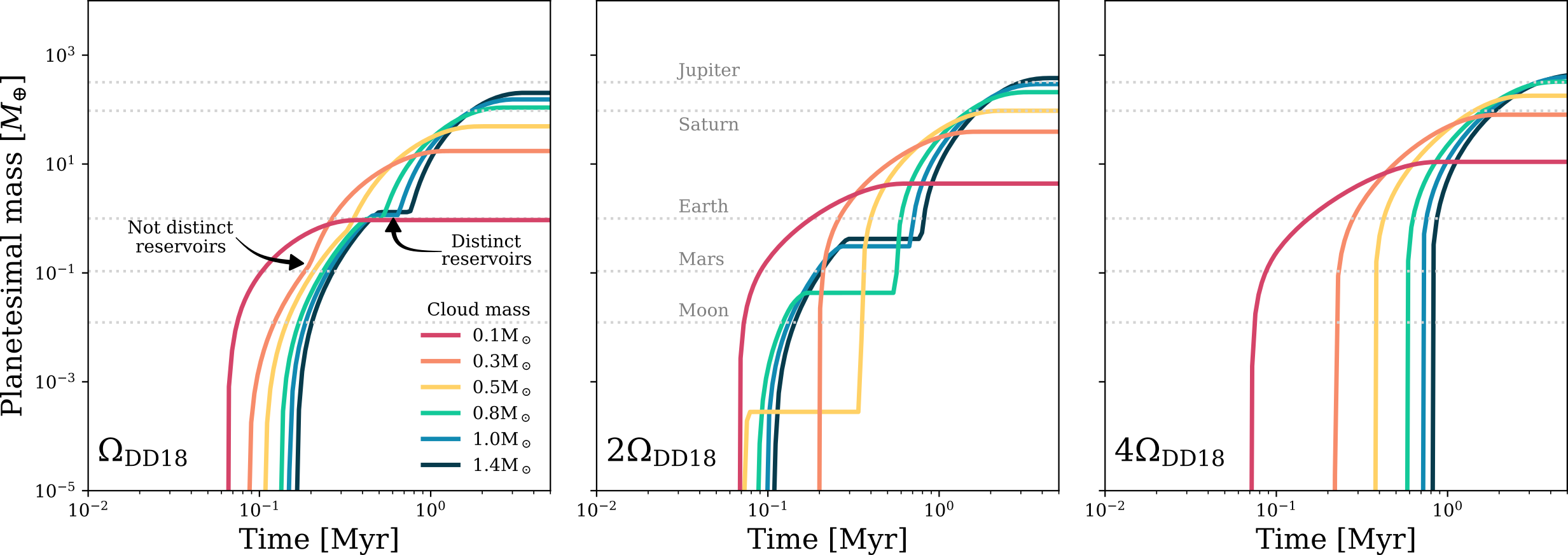}
    \caption{Total mass of formed planetesimals within each simulation as a function of time, shown for different cloud masses (colours) and $\Omega_{\rm{cloud}}$ (labelled in panels). Distinct planetesimal reservoirs are defined when there is a clear break in planetesimal formation (i.e. the mass over time becomes flat) before starting again. The 0.3 and $0.5M_\odot$ cases for $\Omega_{\rm{cloud}}=0.1\Omega_{\rm{DD18}}$ do not have distinct reservoirs, but higher mass clouds do. See section~\ref{subsubsec:planetesimal formation variation} for details.}
    \label{fig:planetesimal mass}
\end{figure*}

\bsp	
\label{lastpage}
\end{document}